\documentstyle[11pt,moriond,epsfig]{article}
\newcommand{\MeV}{{\rm MeV}}

\def\be{\begin{equation}}
\def\ee{\end{equation}}
\def\bea{\begin{eqnarray}}
\def\eea{\end{eqnarray}}

\newcommand{\Nq}{{\mathcal N}_q}
\newcommand{\Ng}{{\mathcal N}_g}

\begin{document}

\null

{\large

\rightline{MPI-PhT/98-37}
\rightline{May 1998}
\vspace{2cm}

 \centerline{\LARGE\bf QCD description of high order factorial moments} 
\vspace{0.2cm} 
 \centerline{\LARGE\bf in quark and gluon jets~\footnote
{\normalsize
 to be published in the Proceedings of the 
XXXIIIrd Rencontres de Moriond, ``QCD and high energy interactions'', 
Les Arcs, France, March 21-28, 1998 } } 

\vspace{1.0cm}

\centerline{SERGIO LUPIA} 
\vspace{1.0cm}

\centerline{\it Max-Planck-Institut f\"ur Physik}
\centerline{\it (Werner-Heisenberg-Institut)}
\centerline{\it F\"ohringer Ring 6, D-80805 M\"unchen, Germany}

\vspace{3.0cm}
\centerline{\bf Abstract}
\bigskip
{ 
\baselineskip=24pt
\noindent
The complete QCD evolution equation for factorial moments
in quark and gluon jets
is numerically solved with initial conditions at threshold
by fully taking into account the energy-momentum conservation law.
Within the picture of Local Parton Hadron Duality, the
perturbative QCD predictions can successfully describe the
available experimental data.
} }

\newpage 
\null \newpage

\vspace*{4cm}
\title{QCD DESCRIPTION OF HIGH ORDER FACTORIAL MOMENTS \\ 
IN QUARK AND GLUON JETS}

\author{SERGIO LUPIA }

\address{Max-Planck-Institut 
f\"ur Physik (Werner-Heisenberg-Institut), \\
F\"ohringer Ring 6, 80805 M\"unchen, Germany}

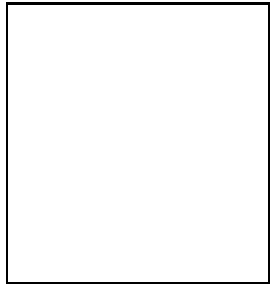
\maketitle\abstracts{
The complete QCD evolution equation for factorial moments 
in quark and gluon jets 
is numerically solved with initial conditions at threshold   
by fully taking into account the energy-momentum conservation law. 
Within the picture of Local Parton Hadron Duality, the 
perturbative QCD predictions can successfully describe the 
available experimental data.} 

\section{Introduction}

An interesting approach 
for describing  multiparticle production in high energy jets 
is provided by the analytical perturbative approach~\cite{dkmt}. 
In this framework, the perturbative evolution is described 
in terms of a parton cascade which is evolved down to small scales
of a few hundred MeV for the transverse momentum,   
and then partonic predictions are compared to hadronic observables, 
according to the notion of local parton hadron duality (LPHD)~\cite{lphd}.
In the description of the partonic cascade QCD coherence is included 
as well as the running of the coupling $\alpha_s$ at one-loop level 
and energy-momentum conservation effects. 
This approach has been shown to be phenomenologically rather successful 
for a large set of observables~\cite{ko}, and 
in particular for jet and particle multiplicity~\cite{lomult}. 
However, recent experimental analyses~\cite{opalmom} have 
found a quantitative discrepancy between data on 
factorial cumulants and factorial moments of the multiplicity
distribution in single quark and gluon jets and theoretical predictions, 
as given in \cite{dremin}, where only the leading
part of the energy-momentum conservation effects have been  
taken into account.

Here we report the result of a new analysis~\cite{lupia}, where 
the complete MLLA evolution
equation of QCD for factorial moments of any order with the full inclusion
of energy-momentum conservation effects have been numerically solved. 
After LPHD, the analytical perturbative approach 
is then shown to be able to describe quantitatively  
the available experimental data on high order factorial moments and
cumulants in single quark and gluon jets. 

\section{Results}

We have considered the complete MLLA evolution equations for the generating
function~\cite{dkmt} and derived the corresponding system of coupled
evolution equations for unnormalized factorial moments $\tilde F^{(q)}$. 
We have then numerically solved this system by imposing boundary conditions
at threshold, i.e., 
$\Ng = \Nq = 1$ and $\tilde F^{(q)}_g = \tilde F^{(q)}_q = 0$ for $q > 1$. 

The three free parameters, i.e., 
the infrared cutoff $Q_0$, the QCD scale
$\Lambda$ and the normalization parameter $K_{all}$, have been kept equal 
to the values fixed in \cite{lomult} to describe the 
jet and particle multiplicities: 
\begin{equation}
K_{all}=1, \qquad \Lambda=500~ \MeV
\qquad  Q_0 = 0.507~\MeV \  .  \label{results}
\end{equation}
A very good quantitative agreement with the experimental data of
\cite{opalmom} is achieved, contrary to previous
comparisons~\cite{opalmom} with approximate predictions~\cite{dremin}. 

\noindent
We refer to ~\cite{lupia} for more details on this new analysis.

\begin{table}     
 \begin{center}
 \vspace{4mm}
 \begin{tabular}{||c|c|c||c|c||}
\hline 
$q$ & \multicolumn{2}{c||}{$F_q^{(q)}$ Quark jet} 
& \multicolumn{2}{c||}{$F_g^{(q)}$ Gluon jet} \\ 
 \hline 
  & Exp~\cite{opalmom} & Theory & Exp~\cite{opalmom} & Theory   \\ 
\hline 
 2 & $1.0820\pm 0.0006\pm 0.0046$ & 1.080 & 
$1.023\pm 0.008 \pm 0.011$  & 1.026 \\ 
 3  & $1.275\pm 0.002\pm 0.017$ & 1.265 & 
$1.071\pm 0.026 \pm 0.034$ & 1.078 \\ 
 4 & $1.627\pm 0.005\pm 0.042$ & 1.600 & 
$1.146\pm 0.059\pm 0.074$ & 1.157  \\ 
 5 & $2.274\pm 0.014\pm 0.093$ & 2.168 & 
$1.25\pm 0.11\pm 0.13$ & 1.268 \\ 
 \hline
 \end{tabular}
 \end{center}
\caption{Experimental data for the  
normalized factorial moments of order $q = 2,\dots,5$ in single
quark and gluon jets of 45.6 and 41.8 GeV respectively are compared 
with our theoretical predictions.} 
\label{tableparameter}
\end{table}

\section{Conclusions}

We have shown that the analytical perturbative approach, based on
perturbative QCD description of the partonic evolution plus Local Parton
Hadron Duality as hadronization prescription, can quantitatively describe
experimental data on multiparticle correlations in single quark and gluon
jets, if the energy-momentum conservation law 
is correctly taken into account.   

\section*{Acknowledgments}
I would like to thank Wolfgang Ochs for many fruitful discussions and
suggestions. 

\end{document}